\documentclass[twocolumn,twoside,slac_two]{revtex4}
\usepackage{graphicx}
\usepackage{fancyhdr}
\begin{document}

\title{ATCA monitoring of gamma-ray loud AGN}

%

\author{J. Stevens}
\affiliation{CSIRO Astronomy and Space Science, Locked Bag 194, Narrabri NSW 2390, Australia}
\author{P.G. Edwards}
\affiliation{CSIRO Astronomy and Space Science, PO Box 76, Epping NSW 1710, Australia}
\author{R. Ojha}
\affiliation{NASA GSFC, Mail Code 661, Greenbelt MD 20771, USA}
\author{M. Kadler}
\affiliation{Universit\"{a}t W\"{u}rzburg, Lehrstuhl f\"{u}r Astronomie, Emil-Fischer-Str. 31 W\"{u}rzburg 97074, Germany}
\author{F. Hungwe}
\affiliation{Rhodes University, P.O. Box 94 Grahamstown 6140, South Africa}
\author{M. Dutka}
\affiliation{Catholic University of America, 620 Michigan Ave., N.E. Washington DC 20064, USA}
\author{S. Tingay, J.-P. Macquart, A. Moin}
\affiliation{Curtin Institute of Radio Astronomy, Curtin University, GPO Box U1987, Perth WA 6845, Australia}
\author{J. Lovell, J. Blanchard}
\affiliation{School of Mathematics \& Physics, University of Tasmania, Private Bag 37, Hobart TAS 7001, Australia}

\begin{abstract}
As a critical part of the Tracking Active Galactic Nuclei with Austral Milliarcsecond 
Interferometry (TANAMI) program \cite{ojha2010}, in November 2007 the Australia
Telescope Compact Array (ATCA) started monitoring the radio spectra of a sample
of southern hemisphere active galactic nuclei (AGN) that were selected as likely candidates
for detection (as well as a control sample) by the Large Area Telescope (LAT)
aboard the \textit{Fermi Gamma Ray Space Observatory}. The initial sample was chosen based on
properties determined from AGN detections by the Energetic Gamma Ray Experiment Telescope
(EGRET). Most of the
initial sample has been detected by \textit{Fermi}/LAT and with the addition of
new detections the sample has grown to include 226 AGN, 133 of which have data for
more than one epoch. For the majority of these
AGN, our monitoring program provides the only dynamic radio spectra available.
The ATCA receiver suite makes it possible to observe several sources at frequencies
between 4.5 and 41 GHz in a few hours, resulting in an excellent measure of spectral
index at each epoch. By examining how the spectral index changes over time, we
aim to investigate the mechanics of radio and $\gamma$-ray emission from AGN jets.
\end{abstract}

\maketitle

\thispagestyle{fancy}


\section{INTRODUCTION}

Results from EGRET revealed a very strong link between $\gamma$-ray emission in AGN and
core-dominated radio emission from relativistic jets, which may originate from super-massive
black hole and accretion disk systems \cite{hartman1999}. 
Recent \textit{Fermi} data have confirmed this \cite{ackermann2011}.
Observations at radio wavelengths were critically important for the understanding of the
$\gamma$-ray emission mechanism in these AGN. In particular, the relative timing of $\gamma$-ray
flares and radio outbursts provides constraints on the mechanisms and location of the flare
origin \cite{jorstad2001}. The main models for $\gamma$-ray emission entail inverse Compton scattering
of low energy photons to high energies by the relativistic electron population in the AGN jets.
The low energy photons could have their origin external to the jet, perhaps in the broad line
region of the AGN \cite{dermer1993} or could originate internal to the jet, being the synchrotron radio/IR/optical
emission from the electrons themselves interacting with the jet magnetic field \cite{bloom1996}.

Radio flux density monitoring observations, which allow for the evolution of the radio emission to be
followed as a function of time, may indirectly reveal when relativistic electrons are being injected
into the AGN jet. Observations at high frequencies, in the optically-thin part of the radio spectrum,
may allow us to probe deep into the structure of the radio jet, to where the $\gamma$-ray emission is
likely generated \cite{lahteenmaki2003}.

\section{ATCA OBSERVATIONS}

Since late 2007, the ATCA has undertaken a project (designated C1730) to regularly observe a number of
southern AGN that have been detected in $\gamma$-rays. For many sources, particularly
those south of $-30^\circ$ declination, this is the only project providing flux density
monitoring data. It measures flux densities for these AGN at frequencies between 4.5 to 41 GHz,
making it ideal for investigating the physics involved in $\gamma$-ray emission.

The ATCA is ideally suited for a program such as C1730. It is a six-element interferometer
located near the town of Narrabri in New South Wales. Each antenna has a 22-m diameter,
with a Cassegrain focus. Each antenna has a suite of cryogenically-cooled receivers, which
cover 67\% of the frequency range between 1.1 - 50 GHz. The telescope can quickly change
between these receivers to make observations at a range of different frequencies within
a short timespan. The C1730 project uses this ability to measure the spectral index of each
source it monitors, by obtaining flux densities at 5.5, 9, 17, 19, 38 and 40 GHz almost
simultaneously. A typical C1730 epoch will observe 5 closely-spaced sources at two simultaneous frequencies in
30 minutes, before switching frequencies and reobserving them again, resulting in full spectral
coverage within 2 hours.

Between the start of the C1730 project in November 2007 and February 2009, observations
were made with the original ATCA correlator, which provided 256 MHz of bandwidth, with
128 MHz at each of two simultaneously measured frequencies. In March 2009, the ATCA correlator
was upgraded to the Compact Array Broadband Backend (CABB, \cite{wilson2011}), which provides 4 GHz of
bandwidth, with 2 GHz at each of two simultaneous frequencies.

The ATCA changes the locations of its antenna several times each year. Five of the six
antenna are set on a 3\,km long East-West track, while the sixth is held stationary 3\,km west
of the western end of this track. This allows for configurations with baseline lengths
between 31m and 6\,km. Antenna can also be placed on a short (214m) North-South track placed
roughly half-way along the East-West track, and this is done when high brightness-sensitivity
is required. The C1730 project has been scheduled in various array configurations, as the
sources that it observes are usually point-like.

Details of the C1730 observing epochs to-date are listed in Table~\ref{epochlist}. The project
commenced with a pilot study before \textit{Fermi} was launched, to trial monitoring of 16
southern AGN detected by EGRET. This successful pilot study paved the way for continued
observations, and with the emergence of the first results from \textit{Fermi} in early 2009, the source list
was quadrupled.

In 2011, the source list was revised based on the first and second \textit{Fermi} catalogs of
detected AGN, and the expanded list of TANAMI targets. The original declination limit of
$-10^\circ$ was changed to better match the TANAMI range ($\delta < -30^\circ$) and reflect the
larger number of southern sources now identified by \textit{Fermi}. The most recent epochs included
150 2LAC sources with $\delta < -40^\circ$, including $\sim30$ proposed counterparts (primarily X-ray
associations) that had not previously been detected at radio frequencies.

\begin{table}[h]
\begin{center}
\caption{C1730 epoch list. For each epoch, the number of sources observed is shown, along
with the ATCA antenna configuration at the time. Details of each configuration can be found
from the ATCA website.}\label{epochlist}
\begin{tabular}{lccl}
\hline
Dates & Array & No. sources \\
\hline
2007 Nov 01/02 & 1.5A & 17 \\
2008 Jan 11/12 & 6A & 20 \\
2008 Mar 14/15 & 1.5D & 20 \\
2008 Oct 13/14 & 6A & 17 \\
2008 Dec 17/18 & 750B & 20 \\
2009 Feb 22/23 & EW352 & 45 \\
2009 Apr 24/25 & H168 & 73 \\
2009 May 18/19 & H214 & 73 \\
2009 Jun 26/27 & H75 & 73 \\
2011 May 17/18 & 1.5B & 50 \\
2011 Aug 31 & 6B & 36 \\
2011 Sep 13 & 6B & 34 \\
2011 Oct 15 & H75 & 44 \\
2011 Nov 02 & 750C & 36 \\
2011 Nov 08/09 & EW367 & 53 \\
2011 Nov 27/28 & 1.5D & 158 \\
2011 Dec 03/04 & 1.5D & 42 \\
\hline
\end{tabular}
\end{center}
\end{table}

Our C1730 observations will contribute significantly toward the identification of the physics of extragalactic
\textit{Fermi} sources, along with monitoring programs at other frequencies, for example:
\begin{itemize}
\item ``Fermi Large Area Telescope Observations of the Active Galaxy 4C +55.17: Steady,
Hard Gamma-Ray Emission and its Implications'' \cite{mcconville2011}
\item ``Fermi Large Area Telescope Observations of Markarian 421: The Missing Piece of its
Spectral Energy Distribution'' \cite{abdo2011d}
\item ``The First Fermi Multifrequency Campaign on BL Lacertae: Characterizing the Low-activity
State of the Eponymous Blazar'' \cite{abdo2011c}
\item ``Insights into the High-Energy Gamma-ray Emission of Markarian 501 from Extensive
Multifrequency Observations in the Fermi Era'' \cite{abdo2011b}
\end{itemize}

C1730 observations are also occassionally made to rapidly follow-up sources that are reported to be
flaring by \textit{Fermi}, for example ATels 3788, 3734, 3713, 3703, 3579 and 3394.
In this way, we have a higher probability of observing
``interesting'' AGN, and thus examining their temporal behaviour during flare events.
With the sensitivity that CABB provides with its 4 GHz of bandwidth, accurate flux density
measurements are possible even for objects that are not normally very bright. For example, in 20 minutes
of on-source integration time, the ATCA with CABB reaches an RMS noise level of 30 uJy at 5.5 GHz, 
40 uJy at 9, 17 and 19 GHz, and 70 uJy at 38 and 40 GHz. It is therefore possible to follow-up
interesting flare events during periods of unscheduled time as small as 2 hours.

\section{BLAZAR VARIABILITY}
All data from C1730 is reduced through the same custom pipeline, ensuring consistency over multiple
epochs. Flux density calibration is achieved via the ATCA primary flux density calibrators PKS B1934-638 --
for frequencies lower than 25 GHz -- and the planet Uranus for higher frequencies.

\begin{figure*}[t]
\centering
\includegraphics[width=135mm,angle=0,scale=1.1]{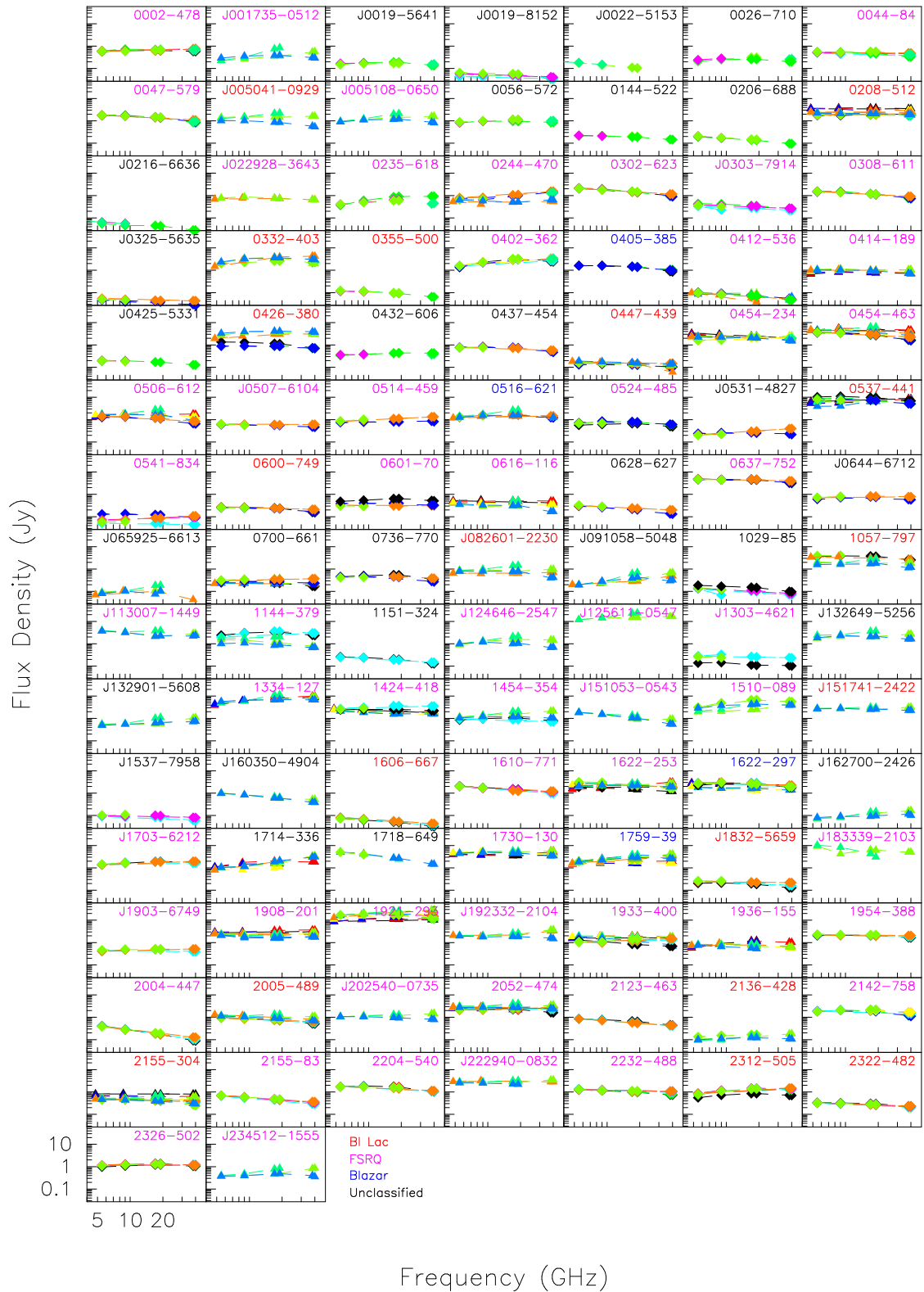}
\caption{Measurements of all sources that have displayed spectral index changes
during the ATCA monitoring campaign. Each epoch is shown as a different
colour/symbol combination. The name of each source is coloured depending on its
classification in the Roma-BZCAT \cite{massaro2009}, CGRaBS \cite{healey2008} and 
1LAC \cite{abdo2010} catalogues,
and the colour legend is shown at the bottom
right of the plot. The frequency and flux density scales are the same for each
plot, and are defined on the lower-left.} \label{allspecvar}
\end{figure*}

Early analysis has focussed on determining which of the AGN monitored by this
program have varied, and how. Using the method described in \cite{angelakis2010}, we have
found 107 sources that have varied between 2007 and now. The spectral curves
for these sources are shown in Figure~\ref{allspecvar}. The variability is characterised by
a change in the spectral turnover frequency. For sources that vary in brightness achromatically,
the spectral turnover frequency will remain constant, but if, for example, the source becomes
disproportionately brighter at higher frequencies, the spectral turnover may shift toward the
higher frequency end.

We used the Roma-BZCAT \cite{massaro2009}, CGRaBS \cite{healey2008} and 1LAC \cite{abdo2010}
catalogues of blazars to separate 
the AGN in this survey into BL Lacs, flat-spectrum radio quasars (FSRQ) and 
unclassified blazars. Sources
that do not appear in any of these catalogues are listed here as ``Unclassified''.
Fifty-nine percent of our sources are blazars, while most others are flat-spectrum
radio sources (FSRS, \cite{healey2007}).

From our survey, 72\% of the BL Lacs have shown spectral-index variability, along with
87\% of the FSRQs and 80\% of the unclassified blazars, while 89\% of the unclassified sources also
show spectral variation. Of these variable sources, 97\% are seen to have a spectral turnover
frequency at the very highest observed frequencies in at least one epoch. This indicates that
the vast majority of our variable sources become periodically brighter at high frequencies,
but do not always stay that way. Further work is required to determine whether these
high-frequency flux increases are associated with $\gamma$-ray activity.

\section{ONGOING WORK}

Observations at the ATCA continue on a regular basis, and analysis of the bulk of the
data has yet to be completed. This section describes some of the analysis we plan to
do in the near future.

The ATCA receivers are capable of measuring polarisation accurately at all the frequencies
observed by this project. It will therefore be possible to examine the polarisation
position angle variability for these blazars, and determine if the magnetic field direction
is changing when the spectral index varies.

The ATCA monitoring data provides information crucial in constructing the 
spectral energy distributions (SEDs) of AGN both
in their flaring and quiescent states. Quasi-simultaneous broadband SEDs are necessary to
distinguish between competing models for the high-energy emission from AGN \cite{boettcher2007,abdo2011}. The
ATCA data are also being used to look for changes in spectral index associated with the
ejection of VLBI components from the AGN cores.

\section{SUMMARY}

The ATCA has monitored a set of 133 AGN over multiple epochs for four years, and will continue
to monitor even more in the future. Through a combination of regular monitoring and
target-of-opportunity observations of flaring AGN, we will investigate the mechanisms
responsible for both radio and $\gamma$-ray emission in their jets.

More information can be found on the project website:
{\tt http://goo.gl/WHxmf}

\bigskip 

\end{document}